\documentstyle[12pt]{article}
\begin{document}

\title
{\Large \bf 
Robustness of Supersensitivity to Small Signals in Nonlinear Dynamical Systems}

\author{ Changsong Zhou$^1$ and  C.-H. Lai$^{1,2}$ \\
        $^1$Department of Computational Science\\
        and $^2$Department of Physics\\
        National University of Singapore,
        Singapore 119260}

\date{}
\maketitle

\begin{center}
\begin{minipage}{14cm}

\centerline{\bf Abstract}
\bigskip

Nonlinear dynamical systems possessing an invariant subspace can display
interesting dynamical behavior, such as on-off intermittency and
bubbling. This letter shows that a class of such systems have  amazing
features of (1)  supersensitivity to small input signals and (2)
 robustness of the supersensitivity in the presence of noise. 
These features make the systems very promising as small signal detectors.

PACS number(s): 05.45.+b;

\end{minipage}
\end{center}

\newpage

Nonlinear dynamical systems with an invariant subspace due to symmetries or
other constraints are of great interest. An example is the synchronization
of  coupled chaotic systems~\cite{yf}.  Such systems can display
interesting and unusual dynamical behaviors, such as on-off
intermittency~\cite{pst} and 
bubbling~\cite{abs}.  In on-off intermittency, the invariant  
manifold is slightly unstable, and the system can remain close to
the invariant manifold for  long periods of  time,  interrupted only
 by some occasional large bursts away from the invariant manifold. 
In bubbling the invariant manifold is stable. However, there are unstable 
invariant sets embedded in the chaotic sets, and small perturbations 
of noise or parameter mismatches can result in large 
intermittent bursts.
This can be harmful in connection with applications of synchronization,
 such as in secure communication~\cite{hgo},  because high-quality
 synchronization is destroyed by bubbling~\cite{abs}.  

The purpose of this letter is to show that such systems are very sensitive to
small constant or time-dependent input signal. With an additional symmetrical
condition, the sensitivity is robust to external noise, which makes the
systems very promising for potential application in weak signal detection.

Inspite of the variety of such systems,  their behavior often can be described
by the following equations:
\begin{eqnarray}
y_{n+1}&=&G(x_n, y_n, a),\\
x_{n+1}&=&F(x_n, y_n, a),  
\end{eqnarray}
where $G(x_n,0,a)=0$,  and the variables $y_n$ and $x_n$ represent the distance from the
invariant manifold $y=0$ and the dynamics within the invariant subspace, respectively.
In general, $y$ and $x$ are vectors. Note that many generic properties of the
above phenomena  can be observed in some very simple systems, and  we consider 
$x$ and $y$  as one-dimensional variables. 
Here $a$ is a parameter which may change the dynamics
within the invariant subspace as well as their stability.
We are interested in the behavior of the system close to the
invariant manifold where the systems can be represented by the
 approximate linear dynamics: 
\begin{eqnarray}
y_{n+1}&=&g(x_n,a)y_n + O(y_n^2), \label{eq1}\\
x_{n+1}&=&f(x_n,a)+O(y_n).
\end{eqnarray}
The nonlinearity  of the systems serves to keep the solution bounded.  
Usually, chaotic signals have quickly (exponentially) decaying correlation. It is
often plausible to assume that the chaotic signals are uncorrelated when 
considering the long time behavior of the system in Eq.~(\ref{eq1}). Based on
this assumption, and for the sake of simplicity and without loss of generality, 
we are led to consider the behavior of the  following simple random driven map
\begin{equation}
y_{n+1}=ax_ny_n + O(y_n^2), \label{eq2}
\end{equation}
where  $x_n$ is a random driving signal. 
Introducing the variable $z_n=\ln|y_n|$, Eq.~(\ref{eq2}) becomes
\begin{equation}
z_{n+1}=z_n+\ln |x_n|+\ln |a|. \label{eq3}
\end{equation}  
The critical value of the  parameter $a$ is defined by $\ln |a_c| + \langle \ln |x_n|\rangle=0$,
where $\langle \cdots \rangle$ represents time average. Close to the critical point
$\delta=(a-a_c)/a_c\ll 1$, Eq.~(\ref{eq3}) can be rewritten as 
\begin{equation}
z_{n+1}=z_n+\delta+\xi_n, \label{eq4} 
\end{equation}
where $\xi_n=\ln |x_n| - \langle \ln |x_n|\rangle $ is a random variable with vanishing mean and
variance $D$. 

To analyze the long time behavior, map~(\ref{eq4}) can be replaced by the corresponding
stochastic differential equation, which is the equation for Brownian motion in
one dimension with a drift. The corresponding Fokker-Planck equation is
\begin{equation}
\frac{\partial W}{\partial t}=-\delta\frac{\partial W}{\partial z}+\frac{D}{2}
\frac{\partial ^2 W}{\partial z^2},
\end{equation}
and the static solution of the probability density $W$ is 
$W(z)=C\exp(\alpha z)$
with $\alpha=2\delta/D$. In the variable $y$, the distribution becomes
$W(y)=C|y|^{\alpha-1}$. 

Now let us  consider that there is a small positive constant  input $p$ 
to the system of Eq.~(\ref{eq2}), namely,
$y_{n+1}=ax_ny_n + O(y_n^2) +p$, 
where $p$ is the order of $10^{-m}, m\gg 1$ (suppose the maximal value of $y_n$ has the order
of unity). For simplicity, we suppose that $y_n>0$ for initial
value $y_0>0$ and $p>0$.
For $y_n\gg p$, the behavior of the system is governed by Eqs.~(\ref{eq2}) and ~(\ref{eq4}).
The effect of the small input  can be regarded as a reflecting barrier to    
the Brownian motion of the system ~(\ref{eq4}), i.e.  $z\geq -m$. On the other hand,
 the state of the system is bounded by the nonlinearity of the system. 
We can introduce a parameter $\tau$ to represent the effect of up-boundary of the
system.   Based on these considerations,
the  behavior of the system with input $p$ 
can be understood by the Brownian motion confined between the two boundaries.
The property of the system is determined by the competition between the 
constant drift $\delta$ and the diffusion $D$. The diffusion is 
dominant for $\delta\sim 0$, i.e. for parameter $a$ close to the critical  
point $a_c$ where the system displays on-off intermittency (bubbling) and  
can both access to the lower and upper boundaries frequently,
becoming sensitive to the small input and producing large bursts. Otherwise,
the drift becomes dominant  when $a$ is far away from the critical point;
for $a<a_c$, the system will spend most of time close to the lower boundary
and produce rare large bursts, and the small input does not lead to
significant large output in the system; for $a>a_c$, the system will spend 
most of time close to the upper boundary and access to the level of the input
rarely, and the small input does not have significant effects on the system
behavior also. We can expect that the system is sensitive
to small input when it is on-off intermittent.   

The above consideration leads to the normalization condition 
$\int_p^{\tau}Cy^{\alpha-1}dy=1$, 
which gives $C=\alpha/(\tau^{\alpha}-p^{\alpha})$.
Now we can estimate the amplitude of the output signals
by the ensemble average
\begin{equation}
\langle y \rangle=C\int_p^{\tau}y^{\alpha}dy
=\frac{\alpha}{1+\alpha}\frac{\tau \beta-p}{\beta-1},
\end{equation}
where $\beta=(\tau/p)^{\alpha}$.  

If $\beta \approx 1$, 
the small input can 
change the behavior of the system greatly. 
For the conditions $|\alpha|\ll 1$, $\tau\gg p$ and $|\alpha| \ln (\tau/p)\ll
1$, one has $\beta \approx 1+\alpha \ln (\tau/p)$, and  
\begin{equation}
\langle y \rangle\approx \frac{\tau}{\ln (\tau/p)}. \label{eq:<y>}
\end{equation}

Eq.~(\ref{eq:<y>}) shows that the average value decreases to zero  with the decrease of
input $p$ only logarithmically, suggesting  that close to the critical point, a very small
input $p$ can produce a relatively large output, i.e., the system  is supersensitive to small
input. A measure of the sensitivity can be  
\begin{equation}
S=\frac{\langle y \rangle}{p}=\frac{\tau}{p\ln (\tau/p)}. \label{eq:s}
\end{equation}
For example, with $\tau=1$ and $p=10^{-15}$, the value of $S$ is about 
$2.9\times 10^{13}$. 

To demonstrate the validity of the above analysis, we carry out simulations with the
following two systems of the form 
$y_{n+1}=ax_nf(y)+p$. For system I, 
$f(y)$ is a piecewise linear map 
\begin{equation}
f(y)=\left\{
\begin{array}{ll}
\frac{c_1}{c_2}(-c_1-c_2-y),& y<-c_1,\\
y,                         & |y|\leq c_1,\\
\frac{c_1}{c_2}(c_1+c_2-y),& y>c_1,\\
\end{array}\right.
\end{equation}
where the parameters $c_1$ and $c_2$ are chosen so that $y_n>0$ for
the positive initial value $y_0$ and $p$, i.e., the bursting in the system  is
symmetry breaking\cite{l}. 
We use $c_1=1$ and $c_2=2$ in our simulations, with   
$x_n$ uniform on $[0,1]$ and thus $a_c=e=2.71828\cdots$ and $D=1$.   
For system II, $f(y)=\sin(y)$, and $x_n$ is a chaotic signal generated by the logistic map 
$x_{n+1}=3.75x_n(1-x_n)$ which gives  $x_n$ a distribution with singularities,
and  $a_c\approx 1.673$ and $D=0.2$. 
$S$ is estimated closed to the critical point for these two  systems.   
With $p=10^{-m}$, $S$ as a function of $m$ is shown in Fig. 1(a). 
The analytical estimation in Eq.~(\ref{eq:s}) with $\tau=1.8$ and $\tau=1.4$ gives good approximation
to the simulation results.
 Fig. 1(b) shows the dependence of $S$ on parameter
deviation $\delta$ from the critical point. The sensitivity is maintained over 
a large range of the parameter $a$.      

The feature of supersensitivity is  maintained even  for time-dependent signals, as for example
\begin{equation}
p_{n+N}=p_n=\left\{ \begin{array}{rl}
p,& 0<n\leq N/2,\\
-p,& N/2<n\leq N.
\end{array}
\right.
\end{equation}
In order for the system to have symmetrical response property to positive and
negative inputs, we require that the map $f(y)$ have odd symmetry $f(-y)=-f(y)$,
 and  on-off intermittency in the system is symmetry-breaking, so that a
positive (negative) small input will eventually lead to only 
 positive (negative) output.  
For $N\gg N_0$, where $N_0$ is the relaxation time of the system, the
sensitivity can be measured by Eq.~(\ref{eq:s}).  

The feature of supersensitivity makes the systems very promising for 
application as sensitive device for small signals.   
In the context of application, we should consider the behavior of the
 system in the presence 
of additive noise, namely,  
$y_{n+1}=ax_nf(y_n)+p_n+e_n$,
where $e_n$ is a small Gaussian white noise with zero mean and standard deviation $\sigma$.
We can study the long time behavior of the system by 
the corresponding stochastic differential equation
$ dy/dt=(\delta+\xi)y+p+e.$
The  Fokker-Planck equation is
\begin{equation}
\frac{\partial W}{\partial t}=-\frac{\partial }{\partial y}\left\{[(\delta+\frac{D}{2})y+p]W\right\}+\frac{1}{2}
\frac{\partial^2}{\partial y^2}[(Dy^2+\sigma^2)W],
\end{equation}
and the static solution under the adiabatic condition $N\gg N_0$ is given by
\begin{equation}
W(y)=C(y^2+\frac{\sigma^2}{D})^{(\alpha-1)/2}\exp[\frac{2p}{\sqrt{D}\sigma}\arctan\frac{\sqrt{D}y}{\sigma}].
\end{equation}
This distribution, however, is very complicated for evaluating $\langle y \rangle$. To simplify the calculation, 
we employ the similar heuristic boundary conditions in the above. Under the conditions $\sigma\ll \sqrt{D}$,
$p\sim \sigma$, we approximate the distribution by 
\begin{equation}
W(y)=\left\{
\begin{array}{ll}
C|y|^{\alpha-1}\exp[\frac{\pi p}{\sqrt{D}\sigma}\hbox{sgn}(y)], & |y|\ge p,\\
0,                                                    & |y|< p.
\end{array}
\right. 
\end{equation}
It is clear that by the limit $\sigma\to 0$, we come back to the result for the noise-free case. 
With this approximation, we obtain that close to the critical point, 
\begin{equation}
\langle y \rangle\approx \frac{\int_{-\tau}^{\tau}yW(y)dy}{\int_{-\tau}^{\tau}W(y)dy}=\frac{\tau}{\ln (\tau/p)}\tanh(\frac{\pi}{\sqrt{D}}R) \label{<y>R},
\end{equation}
where $R=p/\sigma$ provides a natural measure of the signal-to-noise ratio.

The above analysis is  demonstrated by numerical simulations in the presence of noise.  
Fig. 2(a) is a typical response of the system to a noisy small signal and 
Fig. 2(b) shows $\langle y_n \rangle$  for different values of $R$. 
Fig. 2(c) displays the dependence of $\langle y \rangle$ on $R$.  
It is seen that the above approximate analysis gives  a good  account for the results 
in a large range of $R$. 
Over a wide  range of $R$, $\langle y \rangle$ is very close to that of the noise-free case. 
The supersensitivity is thus robust to additive noise. 
This feature of sensitivity is quite different from  that of the sensitivity near the onset
of a period-doubling bifurcation  in many dynamical systems~\cite{wm}. 
There the system is only sensitive
to perturbations near half the fundamental frequency of the system for bifurcation parameter
very close to the onset point. 
 
To examine the performance of the system as a small signal detector,  we  calculate the
probability of bit error $P_b$ in the presence of additive noise. The detection is done by
examining the time average of the output $y_n$ in the duration of a input bit $b_k$, namely
$s_k=1/N\sum_{N(k-1)+1}^{Nk} y_n.$
A bit $b_k$ is detected as $B_k=1\; (-1)$ if $s_k>0 \;(<0)$.   
 For $N\gg N_0$,
the variable $s_k$ is expected to fluctuate around $\langle y \rangle$. Although $y_n$ cannot assumed to be  
uncorrelated, for very large $N$, it might still be plausible to assume that $s_k$ approaches 
a Gaussian distribution with an average $\langle y \rangle$ and a variance $D_N=\Delta/N$, especially in the case that 
$R$ is small and $y_n$ has comparable distribution to positive and negative values. Based  
on this assumption  $P_b$ can be evaluated approximately as
\begin{equation}
P_b=\hbox{Prob}(B_k\neq b_k)=\hbox{Prob}(s_kb_k< 0)\approx \frac{1}{2}[1-\hbox{erf}(\sqrt{\frac{N}{\Delta}}\langle y \rangle)],
\label{p_b}
\end{equation}
where $\langle y \rangle$ is given by Eq.~(\ref{<y>R}).

In our simulations, we estimate $P_b$ with $10^6$ random bits in the system I at $a=2.6$ 
for input levels $p=10^{-4}$ and $p=10^{-6}$. The quantities $N_0$ and $\Delta$ are  estimated in 
simulation with constant input, giving  $N_0=350$, $\Delta=5$ for $p=10^{-4}$ and $N_0=700$, 
$\Delta=4$ for $p=10^{-6}$. Both the results of $P_b$ from  simulations and from estimation in 
 Eq.~(\ref{p_b}) 
are shown in Fig. 3 for $N=3N_0, 5N_0, 10N_0, 15N_0$.  The parameter $\tau$  used to fit 
Eq.~(\ref{p_b}) to the
simulation results       
is $\tau=1.4$ for $p=10^{-4}$ and $\tau=1.1$ for $p=10^{-6}$. It is seen that the estimation can be quite 
good for  $N$ much larger than $N_0$. For $N$ comparable  to $N_0$, the effects of the transient 
process during the relaxation time becomes significant, and the estimation
deviates from the simulation results. For $R$ getting larger, the bursting behavior become
more asymmetrical, and $s_k$ can no longer be approximated  by a Gaussian distribution and the estimation
also deviates  from the simulations results. 

The above results show that detection  error can be quite low even for  a small signal with  a level 
much lower than the environment noise if the signal has a bit duration much larger than the relaxation 
time of the system.  Related to this,   there seems to be a frequency cutoff  above 
which detection becomes  unreliable.     
This frequency gets larger as the input level increases, because the relaxation time becomes
shorter for higher level of input.

The above properties of supersensitivity and its robustness in the presence
of noise is universal in a general class of coupled symmetrical systems
displaying  on-off intermittency with symmetry-breaking. The sensitivity is due
to the  power-law distribution of on-off intermittency of $y$  
in a wide interval
$10^{-m}<|y|<\tau$, so that the system can both access to the level of the
small input and at the same time produce frequent large bursts. 
If the maps are not coupled to random or chaotic driving $x_n$, there is no
diffusion in the system ($D=0$), and the  small input does not have
significant effect on the system output.  For  uncoupled non-on-off maps,
namely, $y_{n+1}=af(y_n)$, if $a<1$ (for the maps in the above), 
the fixed point $y=0$ is stable, and the
small input cannot  produce large output at all; if $a>1$, 
the state $y_n$ can no longer
come to the level of a small input of the order $p=10^{-m} (m>1)$ with significant frequency,
 and the output will not manifest the small input. In both cases, the systems do not 
possess the sensitivity in the coupled, on-off intermittent systems. 
The symmetry-breaking of the bursting also  plays an important role in the
sensitivity  and robustness, because under this condition, a transition of
the state between $y>0$ and $y<0$ is determined only by the switch of the
small signal. If the bursting is not symmetry-breaking, there are additional transitions
between $y>0$ and $y<0$ induced by bursting states, which  will degrade 
the sensitivity and robustness.

In conclusion, we demonstrate 
that a class of nonlinear dynamical systems
having an invariant subspace and displaying on-off intermittency and bubbling have s  the 
feature of supersensitivity to small constant or time-dependent input signals. 
With an additional
odd symmetry condition, the sensitivity is  robust to additive noise. 
The features  make the systems  
very promising for  useful application as  sensitive devices.

\bigskip
{\bf Acknowledgements:}

This work was supported in part by research grant No. RP960689 at the National
University of Singapore.  Zhou is supported by NSTB.

\newpage
{\large \bf Figure Captions}
\begin{description}

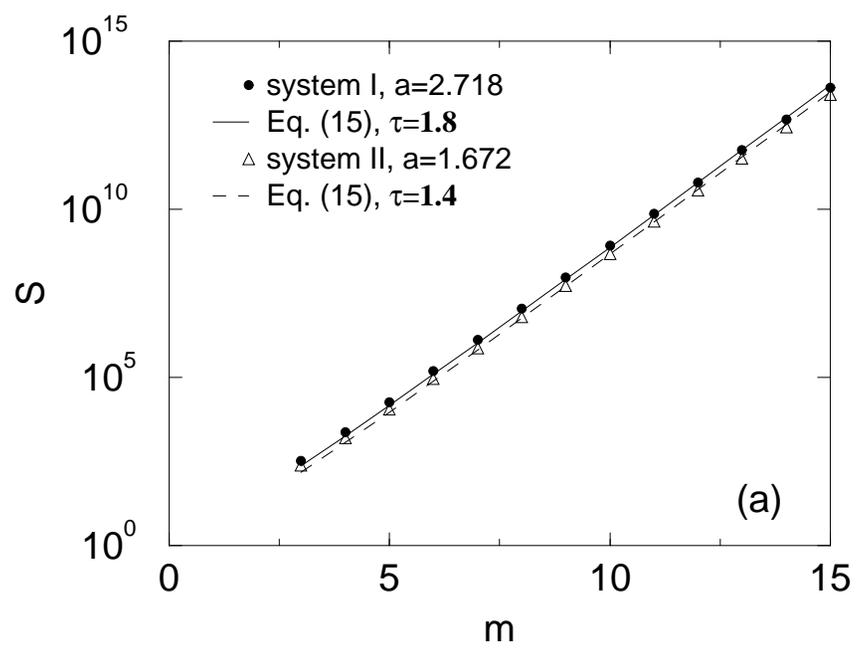
\item Fig. 1. (a) Dependence of the sensitivity $S$ close  to the critical point of the systems 
                 on input $p=10^{-m}$. 
             (b) Dependence of $S$ on $\delta$ for $p=10^{-10}$. 

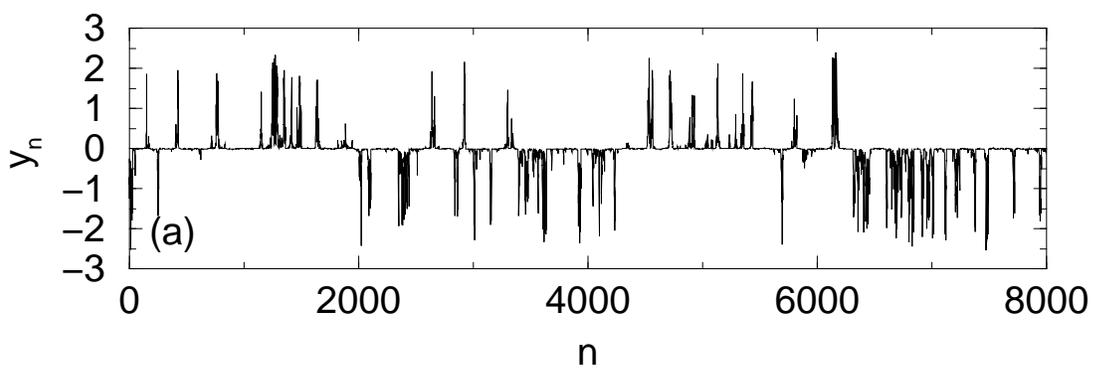
\item Fig. 2. (a) An  example of the bursting behavior of the system  I with $a=2.6$, $N=4000$,
                 $p=10^{-5}$ and $R=0.1$.
               (b)Time series of the ensemble average $\langle y_n \rangle $ over 5000 samples  for 
                the system I with  $a=2.6$, $p=10^{-5}$ and
                $N=4000$. The three plots are: (1) for noise-free case, (2) for $R=0.2$, and
                 (3) for $R=0.05$.
             (c) Ensemble average $\langle y \rangle$ close to the critical point as a function of $R$ for constant
                  input $p=10^{-5}$. The solid lines are estimation of Eq.~(\ref{<y>R}). 

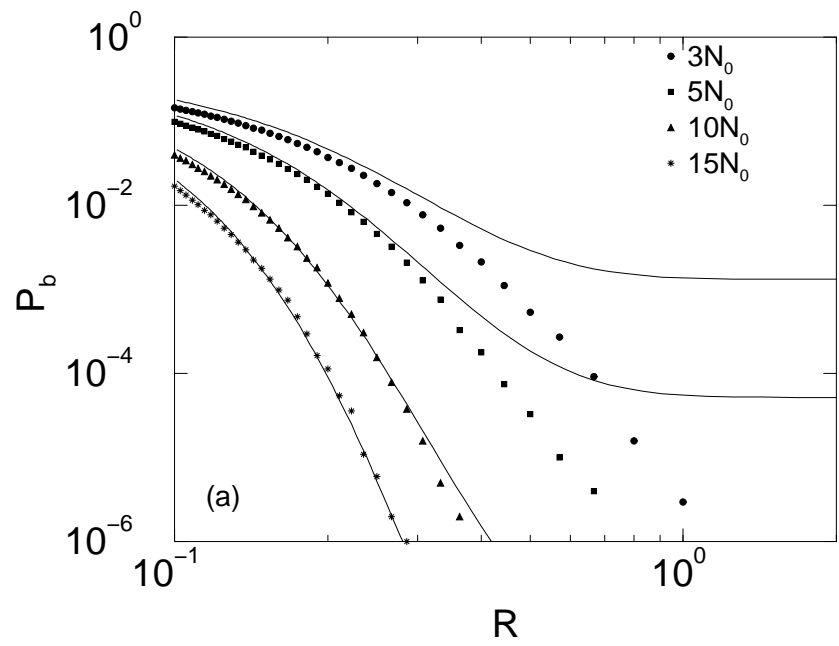
\item Fig. 3. The probability of bit error $P_b$ as a function of $R$ for different levels of input and different
              bit durations.  The solid lines are estimation of Eq.~(\ref{p_b}).  (a) $p=10^{-4}$, $N_0=350$,
$\Delta=5$ and $\tau=1.4$. (b)  $p=10^{-6}$, $N_0=700$,
$\Delta=4$ and $\tau=1.1$ 
                  
\end{description}


\newpage

\begin{thebibliography}{99}
\bibitem{yf} T. Yamada and H. Fujisaka, Prog. Theor. Phys. {\bf 70}, 1240 (1983);
 L. M. Pecora and T. L. Carroll, Phys. Rev. Lett. {\bf 64}, 821 (1990);  
 K. Josic, Phys. Rev. Lett. {\bf 80}, 3053 (1998).              
\bibitem{pst} N. Platt, E. A. Spiegel, and C. Tresser, Phys. Lett. {\bf 70}, 279 (1993);
 N. Platt, S. M. Hammel, and J. F. Heagy, Phys. Rev. Lett. {\bf 72}, 3498 (1994); 
 J. F. Heagy, N. Platt, and S. M. Hammel, Phys. Rev. E {\bf 49}, 1140 (1994);
 Y. H. Yu, K. Kwak, and T. K. Lim, Phys. Lett. A {\bf 198}, 34 (1995);
 A. Cenys, A. Namajunas, A. Tamasevicius, and T. Schneider,
               Phys. Lett. A {\bf 213}, 259 (1996);
 H. L. Yang and E. J. Ding, Phys. Rev. E {\bf 54}, 1361 (1996).
\bibitem{abs} P. Ashwin, J. Buescu, and I. Stewart, Phys. Lett. A {\bf 193}, 126 (1994);
 J. F. Heagy, T. L. Carroll, and L. M. Pecora,
               Phys. Rev. E {\bf 52}, R1253 (1995);
 D. J. Gauthier and J. C. Bienfang, Phys. Rev. Lett. {\bf 77}, 1751 (1996);
 S. C. Venkataramani, B. R. Hunt, E. Ott, D. J. Gauthier, and J. C. Bienfang,
            Phys. Rev. Lett. {\bf 77}, 5361(1996). 
\bibitem{hgo} K. M. Cuomo and A. V. Oppenheim, Phys. Rev. Lett. {\bf 71}, 65 (1993).
\bibitem{l}  Ying-Cheng Lai, Phys. Rev. E {\bf 53}, R4267, (1996).
\bibitem{wm} K. Wiesenfed and B. McNamara, Phys. Rev. Lett. {\bf 55}, 13 (1985).
\end{thebibliography}
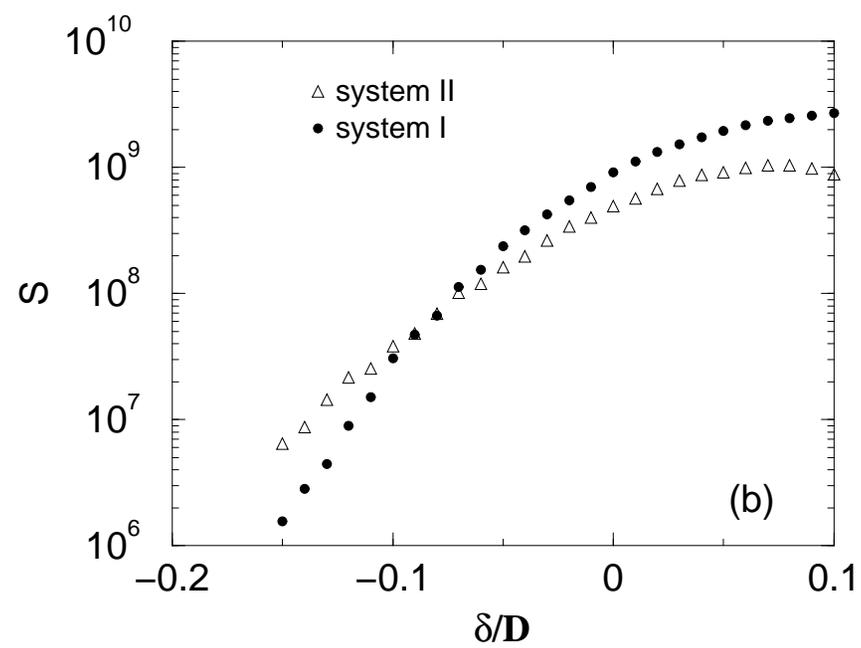
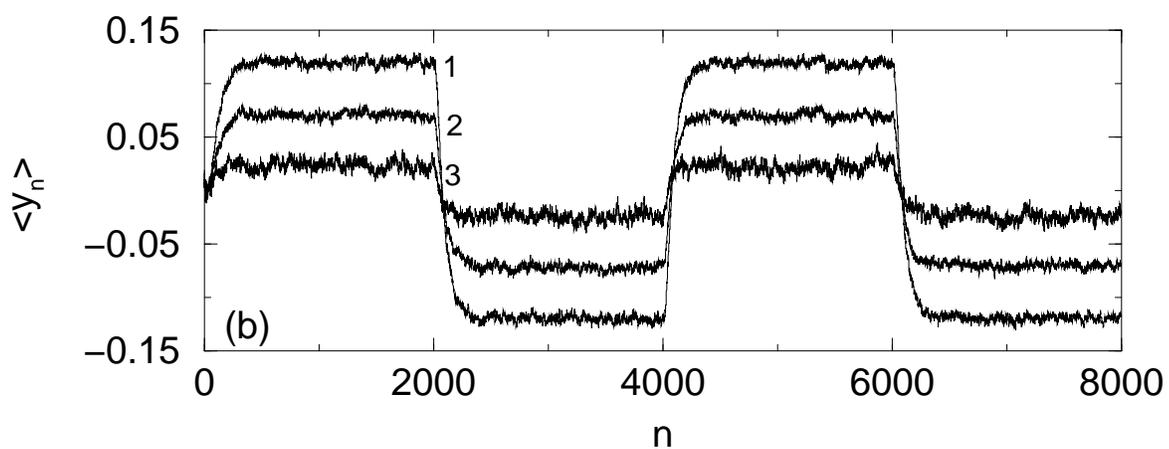
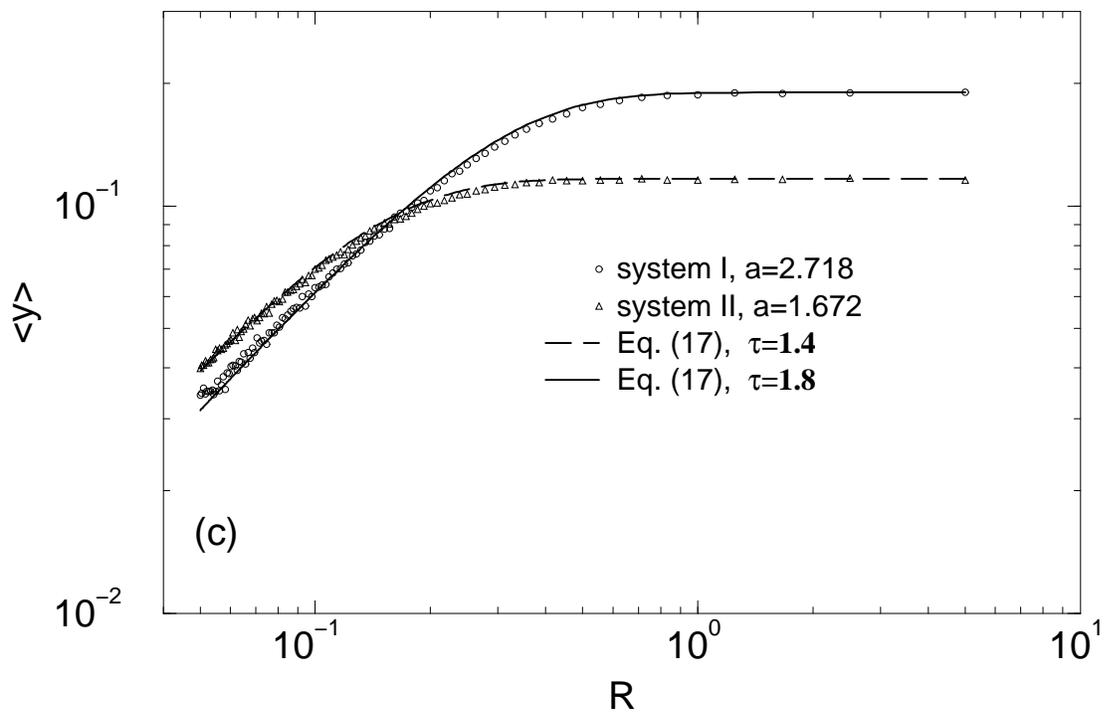
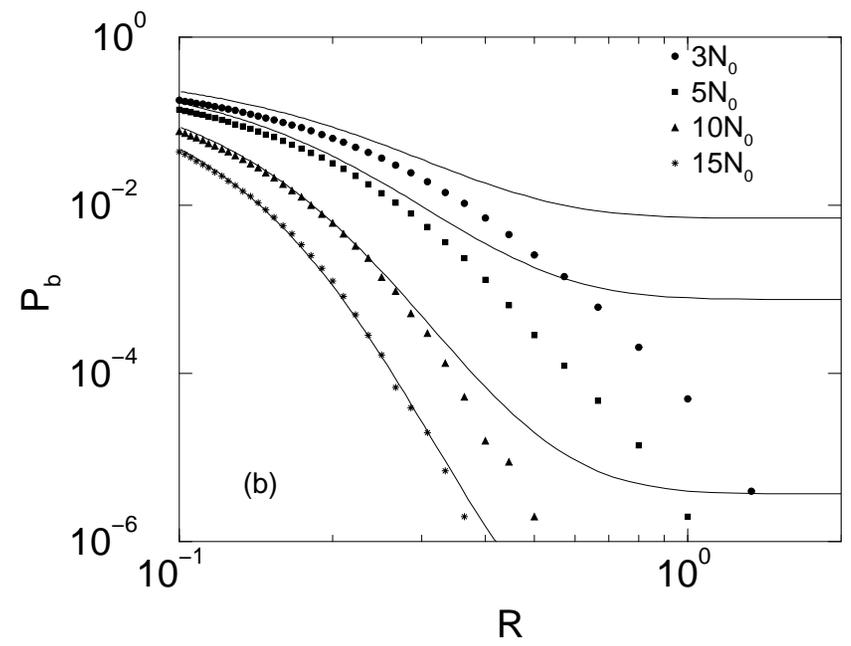
\end{document}